\documentclass[twocolumn]{aastex61}

\usepackage[T1]{fontenc}
\usepackage{amsmath}
\usepackage{booktabs}

\usepackage{savesym}
\savesymbol{tablenum}
\usepackage[separate-uncertainty=true,multi-part-units=single,per-mode=reciprocal]{siunitx}
\restoresymbol{SIX}{tablenum}

\received{May 9, 2017}
\revised{July 3, 2017}
\accepted{July 20, 2017}
\submitjournal{AJ}

\watermark{DRAFT}

\begin{document}

\title{Atmospheric extinction coefficients in the $\mathrm{I_c}$ band
  for several major international observatories:\\Results from the
  BiSON telescopes, 1984 to 2016}

\correspondingauthor{S.~J.~Hale}
\email{s.j.hale@bham.ac.uk}

\author[0000-0002-6402-8382]{S.~J.~Hale}
\author{W.~J.~Chaplin}
\author{G.~R.~Davies}
\author{Y.~P.~Elsworth}
\author[0000-0002-3834-8585]{R.~Howe}
\author{M.~N.~Lund}
\author{E.~Z.~Moxon}
\author{A.~Thomas}

\affiliation{School of Physics and Astronomy, University of
  Birmingham, Edgbaston, Birmingham B15 2TT, United Kingdom}
\affiliation{Stellar Astrophysics Centre, Department of Physics and
  Astronomy, Aarhus University, Ny Munkegade 120, DK-8000 Aarhus C,
  Denmark}

\author{P.~L.~{Pall{\'e}}}
\affiliation{Instituto de Astrof{\'i}sica de Canarias, and Department
  of Astrophysics, Universidad de La Laguna, San Crist{\'o}bal de La
  Laguna, Tenerife, Spain}

\author{E.~J.~{Rhodes},~Jr.}
\affiliation{Department of Physics and Astronomy, University of
  Southern California, Los Angeles, CA 90089, USA}
\affiliation{Astrophysics and Space Sciences Section, Jet Propulsion
  Laboratory California Institute of Technology, 4800 Oak Grove Dr.,
  Pasadena, CA 91109-8099, USA}



\begin{abstract}

Over 30 years of solar data have been acquired by the Birmingham Solar
Oscillations Network (BiSON), an international network of telescopes
used to study oscillations of the Sun. Five of the six BiSON
telescopes are located at major observatories. The observational sites
are, in order of increasing longitude: Mount Wilson (Hale) Observatory
(MWO), California, USA; Las Campanas Observatory (LCO), Chile;
Observatorio del Teide, Iza\~{n}a, Tenerife, Canary Islands; the South
African Astronomical Observatory (SAAO), Sutherland, South Africa;
Carnarvon, Western Australia; and the Paul Wild Observatory, Narrabri,
New South Wales, Australia. The BiSON data may be used to measure
atmospheric extinction coefficients in the $\mathrm{I_c}$ band
(approximately~\SIrange{700}{900}{\nano\metre}), and presented here
are the derived atmospheric extinction coefficients from each site
over the years 1984 to 2016.

\end{abstract}




%
%
%
%

\section{Introduction}
\label{S-Introduction}

The \emph{Birmingham Solar Oscillations Network} (BiSON) is a six-site
ground-based network of solar observatories.  The primary science
output of the network is detection of solar oscillations.  Here, we
take an alternative window into these data and assess the historic
atmospheric column extinction coefficients at each of our
international network sites, over the life of the network.  The
history and performance of the network is detailed
in~\citet{Hale2016}.  In summary, the first instrument was
commissioned at {Observatorio del Teide} in Iza\~{n}a, Tenerife, in
1975, with the additional five nodes coming online throughout the
mid-80s and early-90s.  The observational sites are, in order of
increasing longitude:
{Mount Wilson (Hale) Observatory} (MWO), California, USA;
{Las Campanas Observatory} (LCO), Chile;
{Observatorio del Teide}, Iza\~{n}a, Tenerife, Canary Islands;
{South African Astronomical Observatory} (SAAO), Sutherland, South Africa;
Carnarvon, Western Australia;
{Paul Wild Observatory}, Narrabri, New South Wales, Australia.
The network operates continuously and provides an annual data duty
cycle averaging around~\SI{82}{\percent}.  The locations of the
network nodes are summarised in Table~\ref{table:bisoncoordinates}.

In the next section we will take a brief look at the network
instrumentation and in section~\ref{S-Coefficients} describe how the
atmospheric-extinction coefficients are determined.  We will then go
on to present the historic extinction coefficients of each site in
section~\ref{S-Sites}.

\begin{table*}[t]
    \caption{Coordinates of the six-station Birmingham Solar Oscillations Network (BiSON)}
    \label{table:bisoncoordinates}
    \begin{center}
    \begin{tabular*}{\textwidth}{@{\extracolsep\fill}l r r r r}

    \toprule

    Location & Longitude & Latitude & Altitude & Commissioned\\
             &   [deg E] &  [deg N] &      [m] &       [year]\\

    \midrule
    
    Mount Wilson, California, USA        & -118.08 & +34.13 & 1742 & 1992\\
    Las Campanas, Chile                  &  -70.70 & -29.02 & 2282 & 1991\\
    Iza\~{n}a, Tenerife, Canary Islands  &  -16.50 & +28.30 & 2368 & 1975\\
    Sutherland, South Africa             &  +20.82 & -32.38 & 1771 & 1990\\
    Carnarvon, Western Australia         & +113.75 & -24.85 &   10 & 1985\\
    Narrabri, New South Wales, Australia & +149.57 & -30.32 &  217 & 1992\\

    \bottomrule

    \end{tabular*}
    \end{center}
\end{table*}

%
%
%
%

\section{Instrumentation}
     \label{S-Instrumentation}

The BiSON solar spectrometers provide very precise measures of the
disc-averaged line-of-sight velocity of the solar surface.  This is
done by comparing the wavelength of the potassium absorption line
at~\SI{769.898}{\nano\metre} formed within the Sun, with the same line
in a vapor reference cell on Earth.  The spectrometers typically have
three photo-detectors.  Two of the detectors measure the intensity of
the light scattered from the vapor cell, and the third measures the
intensity of light transmitted directly through the instrument.  The
light is pre-filtered using an $\mathrm{I_c}$ band filter
(approximately~\SIrange{700}{900}{\nano\metre}) formed from Schott~RG9
and~KG4 glass, and the bandwidth is then reduced again
to~\SI{15}{\angstrom} using an interference filter centred
on~\SI{769.9}{\nano\metre}.  The width of the potassium absorption
line is significantly narrower than~\SI{15}{\angstrom}, and so the
measurement of the transmitted light can be considered to be a
measurement of the direct-Sun radiance near the center of the
$\mathrm{I_c}$ band -- essentially the instrument becomes an automated
solar photometer.  Where a measurement of the transmitted light is not
available, the light scattered from the vapor cell can also be used as
a proxy for the transmitted intensity -- see the appendix for further
details.  Figure~\ref{fig:sutherland_intensity} shows a typical day of
data captured from the site at Sutherland, along with the variation in
airmass during the day.  The data have been pre-processed to remove
any periods of instrumental failure and cloudy conditions.  Even thin
cirrus produces an easily identifiable reduction in data quality, and
so the data analysed in this paper are from clear sky observations
only.
     
\begin{figure}
\plotone{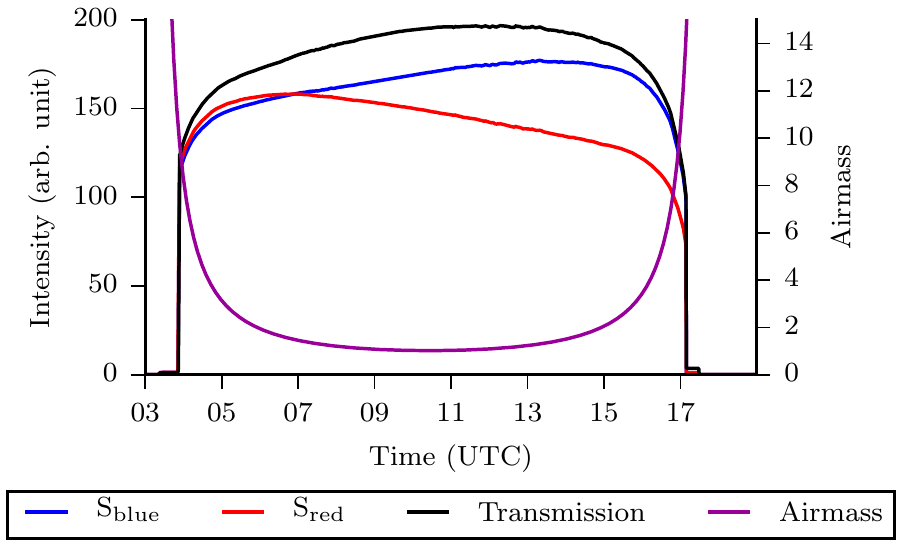}
\caption{Data from 2014~December~1 at Sutherland, South Africa.  The
  black line shows the transmitted solar intensity.  The blue and red
  lines show respectively the measured intensity at points on the
  shorter-wavelength ($\mathrm{S_{blue}}$) and longer-wavelength
  ($\mathrm{S_{red}}$) `wings' of the potassium absorption line
  at~\SI{769.898}{\nano\metre}.  The variation in airmass is shown by
  the magenta line, measured against the
  right-axis. \label{fig:sutherland_intensity}}
\end{figure}

\section{Deriving extinction coefficients}
     \label{S-Coefficients}
     
\begin{figure*}
\plotone{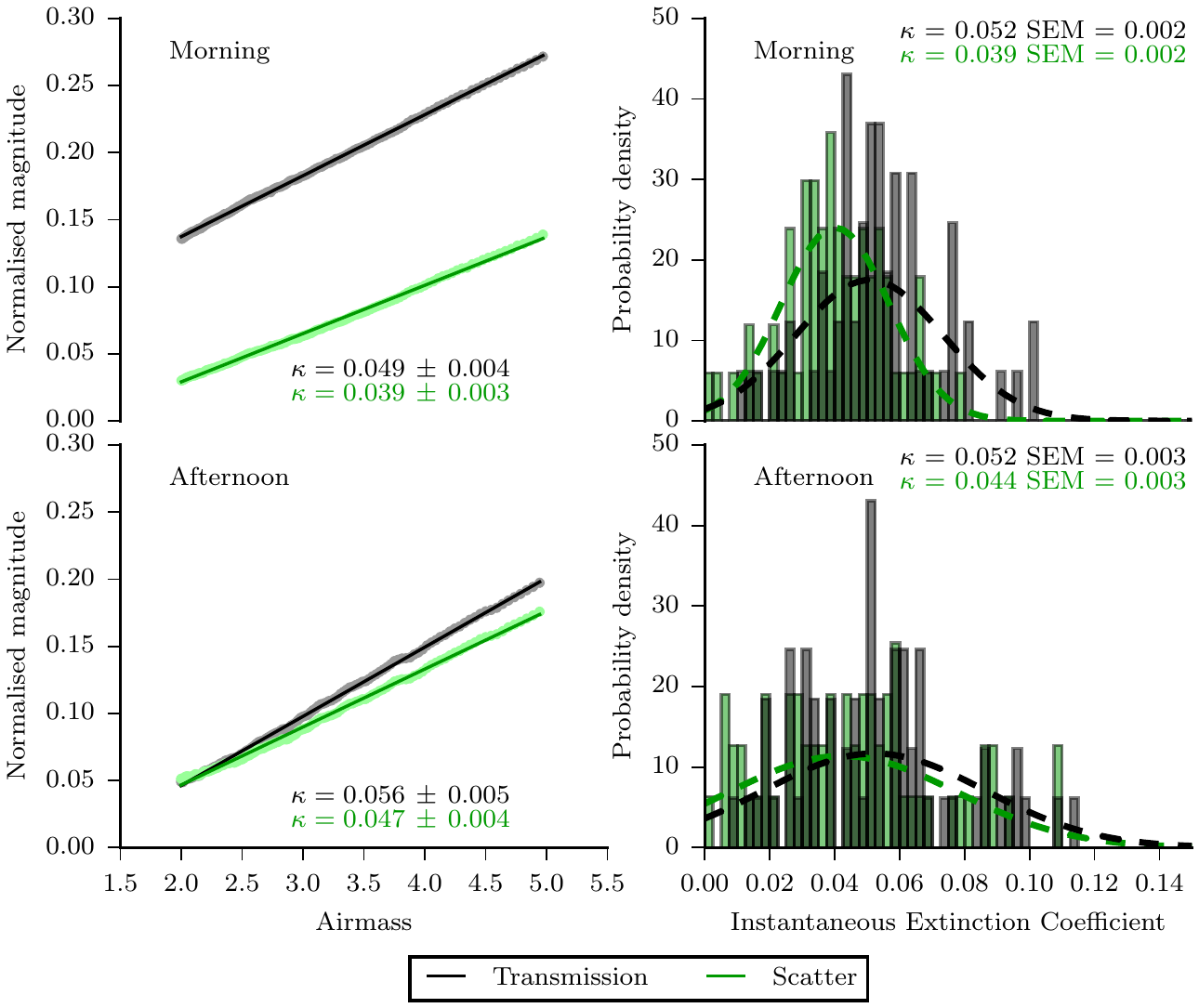}
\caption{Morning and afternoon extinction coefficients for
  2014~December~1 at Sutherland, South Africa.  The values determined
  are shown in the plot.  Left column: The formal uncertainty on the
  fit is calculated from the covariance matrix returned by the fitting
  function.  Right column: The uncertainty on the histogram is the
  standard-error of the median (SEM), i.e., 0.741 times the
  interquartile range of the distribution divided by the square-root
  of the number of points.  The dashed lines on the histograms
  represent for comparison the equivalent Gaussian profile for the
  measured mean and standard
  deviation. \label{fig:extinction_example}}
\end{figure*}

Atmospheric extinction has three main components: Rayleigh scattering,
scattering due to aerosols, and molecular absorption.  The strongest
absorption effects are due to molecular oxygen and ozone which both
absorb in the ultraviolet, and water vapor which absorbs in the
infrared.  At the {BiSON} observational wavelength
of~\SI{769.9}{\nano\metre}, Rayleigh scattering is at a level of a few
percent, and there is no molecular absorption: the observed
atmospheric extinction is dominated by the contribution of aerosols.

The Beer-Lambert law states that the transmittance, $T$, of a material
is related to its optical depth, $\tau_\lambda$, by
\begin{equation}\label{eq:atmos_ext1}
  T = \frac{I}{I_0} = e^{-\tau_{\lambda} A} ~,
\end{equation}
where $I_0$ is the solar extraterrestrial radiance (i.e., at zero
airmass), and $I$ is the direct-Sun radiance.  In this case
$\tau_\lambda$ is the column atmospheric aerosol optical depth (AOD)
per unit airmass, and $A$ is the relative optical airmass as a
function of solar zenith angle.  The aerosol optical depth is
typically quoted as unitless when considering only the unit airmass at
the zenith.  By taking the natural logarithm of both sides we obtain,
\begin{equation}\label{eq:atmos_ext2}
  \ln(I/I_0) = -\tau_{\lambda} A ~,
\end{equation}
which gives a convenient linear relationship where the gradient of the
relationship is a measure of $\tau_\lambda$, and $I_0$ is now simply a
normalization factor taken as the maximum intensity measured on a
given day.  For astronomical use, we rescale AOD in terms of
magnitude,
\begin{equation}\label{eq:atmos_ext3}
\begin{split}
  \kappa_\lambda &= -2.5 \log_{10}(e) (-\tau_\lambda)  ~,\\
                &= 1.086 \tau_\lambda                 ~,
\end{split}
\end{equation}
where $\kappa_\lambda$ is the atmospheric extinction coefficient, with
units of magnitudes per airmass. More accurately, this is the column
extinction coefficient since we do not include any knowledge on the
vertical structure of the atmosphere.  Readers in the climate modeling
and aerosol communities should divide the values presented here by
\num{1.086} in order to recover the total column-aerosol in terms of
aerosol optical depth (AOD).

In this analysis, the known zenith-angle was used to calculate the
airmass based on~\citet{Kasten:89} who define the airmass as,
\begin{equation}\label{eq:kasten}
  A = \frac{1}{\cos{z} + 0.50572 (6.07995 + 90 - z)^{-1.6364}} ~,
\end{equation}
where the zenith-angle $z$ is in degrees.  This model gives an airmass
of approximately~38 at the horizon, producing good results for the
whole range of zenith-angles.

The extinction coefficients were determined using two methods: Firstly
by making a standard linear least-squares fit of the magnitude-like
value from equation~\ref{eq:atmos_ext2} against airmass, and secondly
by calculating the non-overlapping independent first-differences and
then obtaining statistical estimations from the histogram of a
timeseries of,
\begin{equation}\label{eq:firstdifference}
  \frac{\mathrm{d}\,m}{\mathrm{d}\,A} = \frac{m_i - m_{i-1}}{A_i - A_{i-1}} ~,
\end{equation}
where $m$ is the magnitude-like value from
equation~\ref{eq:atmos_ext2}, $A$ is airmass, and $i$ is the sample
index incremented in steps of two.  In our fits we consider only air
masses in the range of 2~to~5, corresponding to zenith-angles between
approximately \SIrange{60}{80}{\degree}, since this is the region
where airmass is changing most linearly.  Below two air-masses the
change does not follow a strictly linear relationship and is not well
described by a straight line fit.  The rate-of-change is also too low
to allow good fitting.  Additionally, we need to ensure we remove the
seasonal variations in minimum air-mass due to the changing maximum
altitude of the Sun throughout the year, since this could introduce an
artificial seasonal effect in the derived extinction values.  Above
five air-masses the rate of change is too high, producing differential
extinction across the extended source of the
Sun~\citep{doi:10.1093/mnras/stu803}.  The pre-meridian and
post-meridian values (hereafter referred to as `morning' and
`afternoon') are fitted separately, since it is expected that these
will differ due to local environmental considerations.  The results
from both techniques for the same day as in
Figure~\ref{fig:sutherland_intensity} are shown in
Figure~\ref{fig:extinction_example}.

The coefficient estimation technique and the selection of airmass
range-limits affects the value of the determined extinction.  In order
to investigate the robustness of our parameters, a randomization trial
was performed where rather than fixing the lower and upper airmass
limits at 2 and 5 respectively, they were randomly selected each day
between 2--3 and 4--5 air-masses.  Five realisations were then
generated for the full timeseries of fitted extinction gradients from
each site, and the absolute difference was compared between each
realisation and the gradients measured when the airmass limits were
fixed.  A similar comparison was made between the timeseries of fitted
gradients, and a timeseries of median first-differences.  In all
cases, the mean difference was less than~\SI{4}{\percent} of the mean
extinction.  The standard deviation of the difference was 3~to~6~times
lower than the measured extinction standard deviation.  Any systematic
offsets or increase in scattering due to the processing techniques are
at a level significantly lower than that due to real physical effects.
These two techniques are considered to be equivalent, and the results
presented here are produced solely from the first method using linear
least-squares fitting.

In the next section we present fitted extinction coefficients for all
the historic data from each BiSON site.  For clarity the units of
extinction will no longer be stated on each value in the text.  All
extinction values are specified in magnitudes per airmass.  During the
discussion for each site, we quote either mean, mode, or median values
depending on the values required for comparisons with other studies.
For consistency, a summary of the coefficients is given in
Table~\ref{table:extinction_coefficients}.  Since there are
significant seasonal differences, we also present the values measured
over two months for each mid-summer (July-August in the northern
hemisphere, January-February in the southern hemisphere) in
Table~\ref{table:summer_extinction_coefficients}, and during
mid-winter (January-February in the northern hemisphere, July-August
in the southern hemisphere) in
Table~\ref{table:winter_extinction_coefficients}.

%
%
%
%

\section{Sites}
\label{S-Sites}

\subsection{Iza{\~n}a, Tenerife}

\begin{figure*}
\plotone{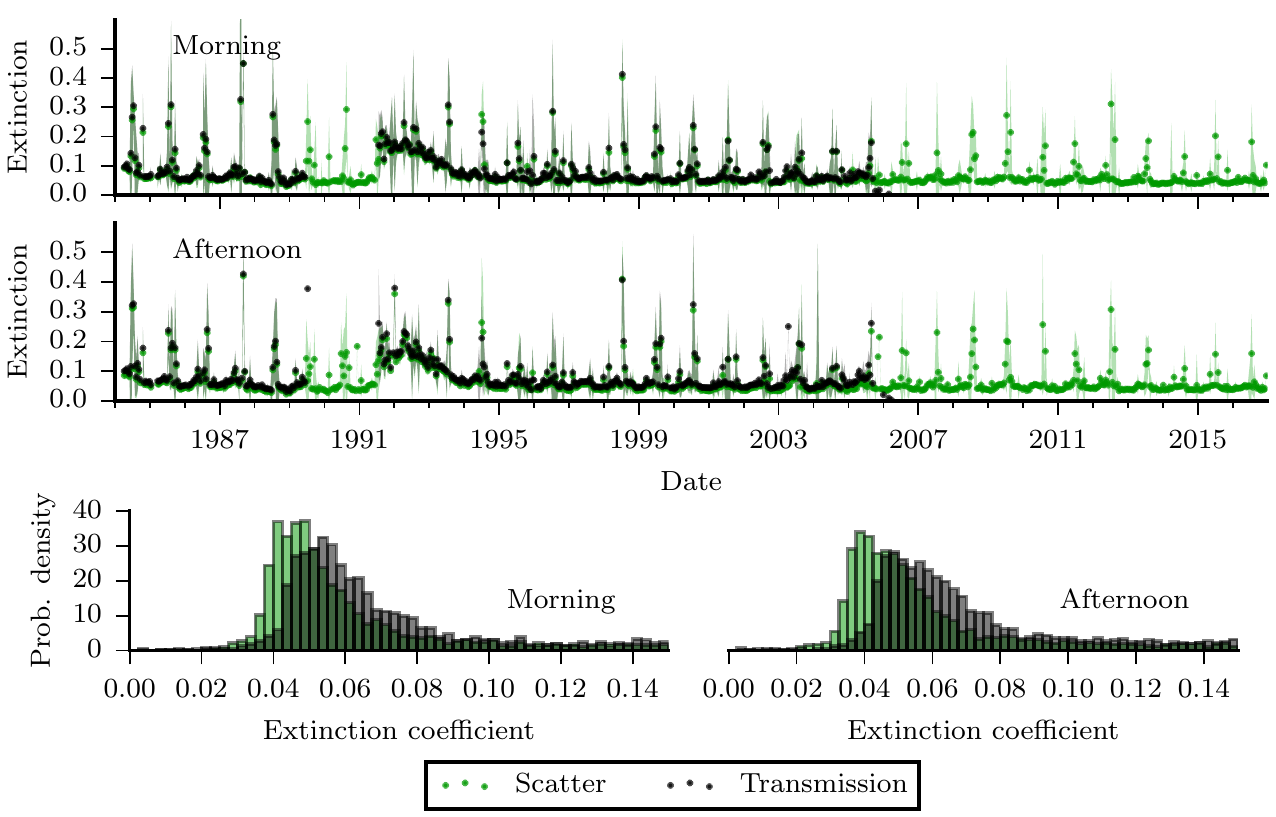}
\caption{Extinction coefficients and statistical distribution from
  Iza{\~n}a, Tenerife.  Each dot represents the median value over
  14~days.  The coloured banding represents $\pm3$ times the
  standard-error on each median. Dates indicate January~1 for each
  year.}\label{fig:izana_extinction}
\end{figure*}

The BiSON node at Tenerife~\citep{2014MNRAS.443.1837R} is based at the
Observatorio del Teide, which is operated by the IAC (\emph{Instituto
  de Astrof{\`i}sica de Canarias}).  The Canary Islands are located
about~\SI{100}{\kilo\metre} to the west of the North African coast.
The islands are close to the Western Sahara and so during the summer
months they frequently experience high concentrations of mineral dust
in the atmosphere.  This aerosol concentration is easily seen in
Figure~\ref{fig:izana_extinction} as the strong seasonal variation in
extinction.

The modal values of the extinction distributions are~0.054 in the
morning and~0.046 in the afternoon for transmission.  The standard
deviation on these values is~0.09, the high value being indicative of
the large scatter in values between summer and winter periods.  The
difference in the values pre-meridian and post-meridian is likely due
to the surrounding geography, where morning extinction effects are
through atmosphere over North Africa and afternoon is over the
potentially clearer North Atlantic ocean.  If only the winter months
are considered, then the modal value for both morning and afternoon
drops to~0.045 with correspondingly reduced scatter.  Mineral dust
events are typical between June and October, where extinction values
anywhere from~\SIrange{0.1}{0.8}{} may be experienced.  The extinction
values derived from the scattered-light are a slight underestimate as
expected (see appendix), but otherwise show the same trends.  The
distribution of extinction coefficients appears to show two combined
trends, the first a set of normally-distributed values centred on
approximately~\num{0.05}, and a long positive tail that corresponds to
the periods of mineral dust events.  If the summer and winter periods
are analysed separately, then the winter does indeed show a mean of
approximately~\SIrange{0.05}{0.06}{}, and the summer a much higher
mean of~\SIrange{0.14}{0.15}{} with correspondingly higher standard
deviation.  A similar atmospheric analysis on data from telescopes on
the Canary Archipelago, which included our BiSON data from this
instrument, was made by~\citet{doi:10.1175/JCLI-D-14-00600.1} and the
modal values are consistent given the uncertainties.
\citet{doi:10.1175/JCLI-D-14-00600.1} provides a thorough
investigation into the occurrence of dust events, and how they change
over both short seasonal periods and longer time scales.  Several
authors have investigated this in more detail \{see, e.g.,
\citet{1998NewAR..42..529G, 1998A&AS..129..413J, 2004SPIE.5489..138S,
  1538-3873-122-895-1109, QJ:QJ2170}\}.

\citet{Siher2002} present extinction values for the IRIS
(\emph{International Research of the Interior of the Sun}) site based
at Iza{\~n}a.  IRIS was a similar network to BiSON and used a similar
observational technique but made use of the shorter wavelength sodium
absorption line at~\SI{589.6}{\nano\metre}.  They quote an average
extinction value of~\num{0.111}, which is slightly higher than the
mean values found here.  The higher value is expected due to their use
of the shorter wavelength, since shorter wavelengths tend to suffer
greater extinction.  \citet{1998A&AS..129..413J} found the extinction
to vary during~1984 to~1989 between~\SIrange{0.04}{0.07}{}
at~\SI{680}{\nano\metre}, which is in agreement with the values found
here for Iza{\~n}a.  During dust storms \citet{1998A&AS..129..413J}
reported values up to~\num{0.8}, which again is in agreement with our
findings.  The GONG (\emph{Global Oscillation Network Group})
site-survey~\citep{1994SoPh..152..351H} at Iza{\~n}a measured an
average extinction value of~\num{0.1169} from 1985~September to
1993~July.  GONG is another network similar to BiSON.  Their initial
site-survey used a normal incidence pyrheliometer (NIP) manufactured
by Eppley Laboratories~\citep{1986SoPh..103...33F} which has a wide
spectral sensitivity range of~\SIrange{250}{3000}{\nano\metre}.  Light
from the Sun is broadly like that of a black body at a temperature
of~\SI{6000}{\kelvin}, meaning that the intensity peaks at a
wavelength of approximately~\SI{500}{\nano\metre} and decays quickly
in the infrared.  The value measured by the pyrheliometer will be
strongly weighted towards the peak wavelength of the solar spectrum.
\citet{tn031} discusses the wavelength dependence of typical expected
atmospheric extinction from~\SIrange{300}{1100}{\nano\metre} at the
Roque de los Muchachos Observatory, on the adjacent island of La
Palma.  At~\SI{500}{\nano\metre} an extinction coefficient
of~\num{0.1244} can be expected, very close to the value determined
during the {GONG} site-survey.  At~\SI{300}{\nano\metre}, near the
short end of the pyrheliometer sensitivity range, values well
over~\num{3} can be expected.  The average extinction measured by the
pyrheliometer will be much higher than the monochromatic values
measured by {BiSON} at~\SI{769.9}{\nano\metre}, and unfortunately this
means that no comparison can be made between the results from {BiSON}
and the {GONG} site-survey.

Probably the most striking feature in
Figure~\ref{fig:izana_extinction} is the increase in extinction
following the eruption of Mount Pinatubo in the Philippines
on~1991~June~15.  There is also a hint of the tail-end of effects from
the El~Chich{\'o}n eruption in Mexico in~1982~April where the start of
the data in~1984 show extinction values around~\num{0.1}, double the
typical value expected outside of a dust intrusion.
Both~\citet{1998NewAR..42..529G} and~\citet{1538-3873-122-895-1109}
have previously observed these features from telescopes based at the
Canary Islands.

\subsection{Carnarvon}

\begin{figure*}
\plotone{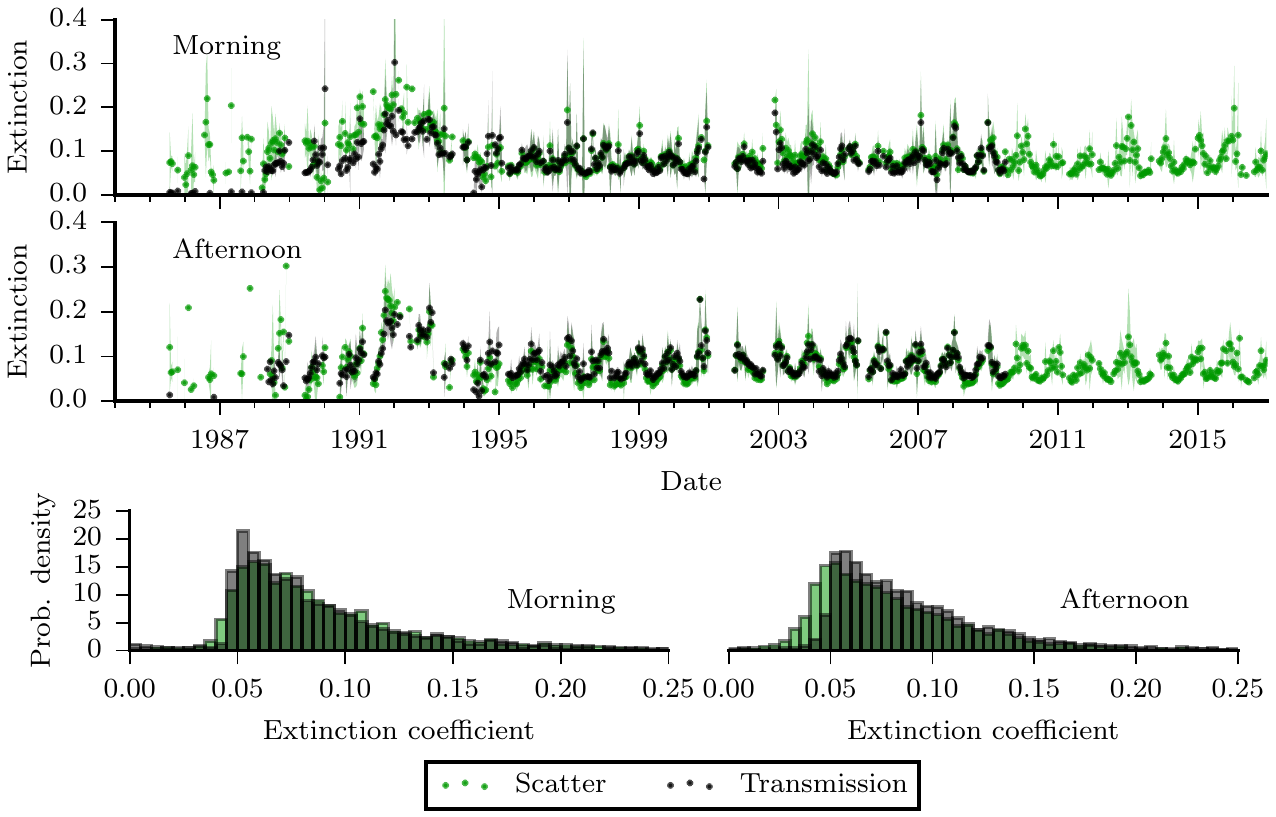}
\caption{Extinction coefficients and statistical distribution from
  Carnarvon, Western Australia.  Each dot represents the median value
  over 14~days.  The coloured banding represents $\pm3$ times the
  standard-error on each median. Dates indicate January~1 for each
  year.\label{fig:carnarvon_extinction}}
\end{figure*}

The BiSON node at Carnarvon is based at the historic \emph{Overseas
  Telecommunications Commission Satellite Earth Station},
around~\SI{900}{\kilo\meter} north of Perth in Western Australia.  The
measured extinction coefficients over the operational lifetime of the
site are shown in Figure~\ref{fig:carnarvon_extinction}.

The median morning extinction is~\num{0.072}, and the afternoon
is~\num{0.078}.  This increases to approximately~\num{0.1} in the
summer and decreases to around~\num{0.06} in the winter.  The standard
deviation shows a similar increase in summer compared to winter.  In
the summer the standard deviation is noticeably lower in the afternoon
compared to the morning, at~\num{0.05} and~\num{0.07} respectively.
Morning data are collected over the plains of Western Australia, and
afternoon data are through air over the Indian ocean, and so it is
expected that the sandy environment of Carnarvon would have greater
impact on extinction during morning observations.  There are no
comparison sites available near Carnarvon.

\subsection{Sutherland}

\begin{figure*}
\plotone{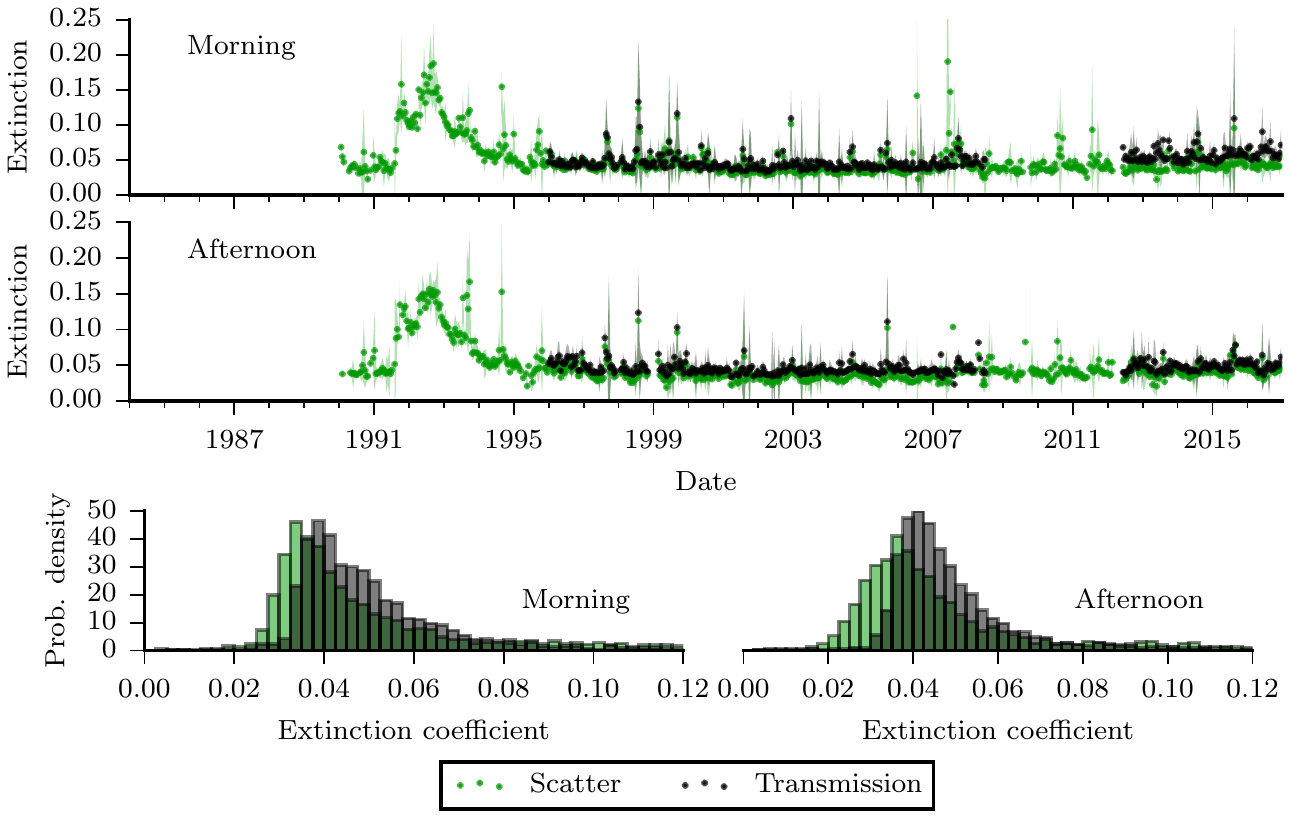}
\caption{Extinction coefficients and statistical distribution from
  Sutherland, South Africa.  Each dot represents the median value over
  14~days.  The coloured banding represents $\pm3$ times the
  standard-error on each median. Dates indicate January~1 for each
  year.\label{fig:sutherland_extinction}}
\end{figure*}

The BiSON node at Sutherland is situated~\SI{360}{\kilo\meter}
north-east of Cape Town, at the \emph{South African Astronomical
  Observatory} (SAAO).  The measured extinction coefficients over the
operational lifetime of the site are shown in
Figure~\ref{fig:sutherland_extinction}.

The median morning extinction is~\num{0.046}, and afternoon
is~\num{0.045}.  There are no significant differences in environment
around Sutherland, and this is reflected by the stability of the
extinction coefficients between morning and afternoon.  The site shows
stable performance with little change in the median values throughout
a year.  The atmosphere is particularly stable, with standard
deviations of between~\num{0.04} and~\num{0.06} in the winter, and
approximately~\num{0.02} during the summer.
\citet{1995Obs...115...25K} state a mean extinction coefficient
of~\num{0.07} in the~$\mathrm{I_c}$ band which is slightly higher than
found here, but is within our measured standard deviation.

\subsection{Las Campanas}

\begin{figure*}
\plotone{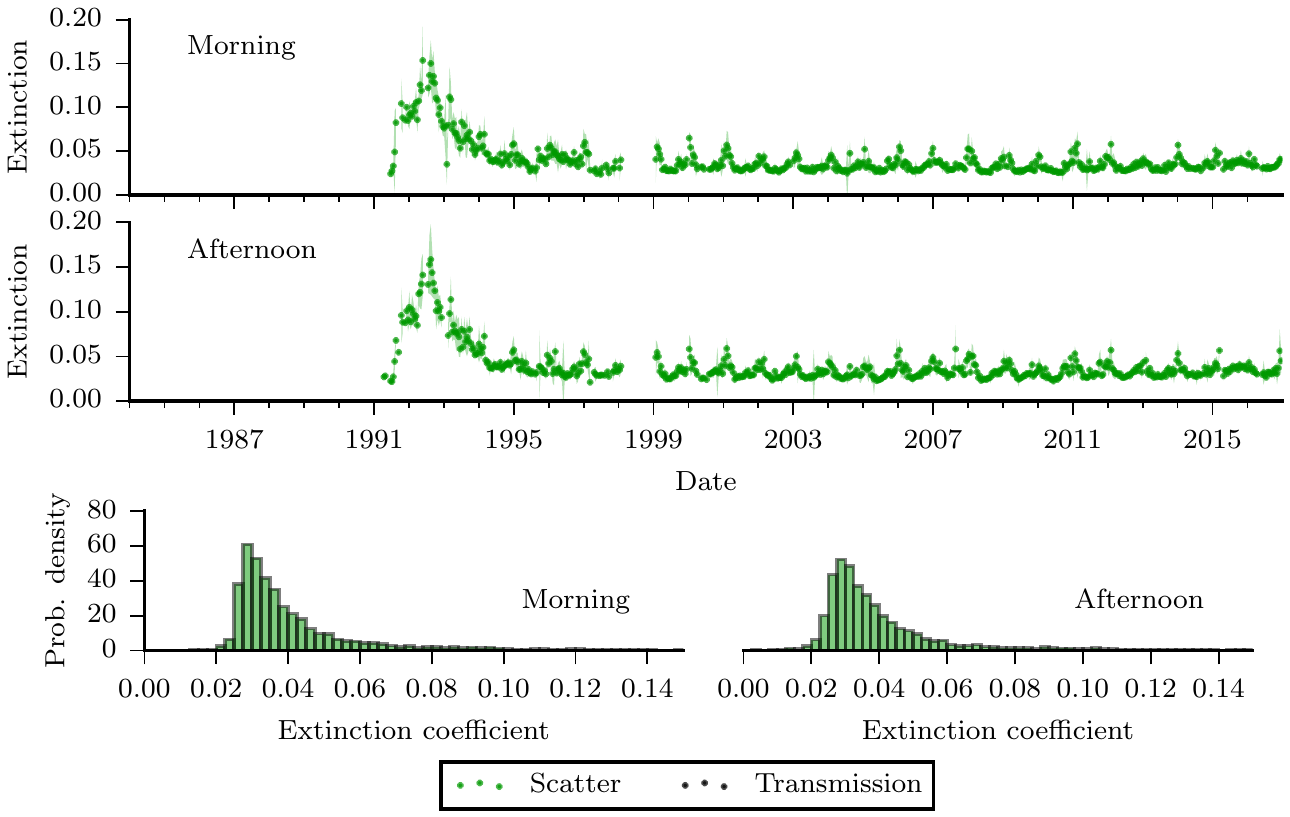}
\caption{Extinction coefficients and statistical distribution from Las
  Campanas, Chile.  Each dot represents the median value over 14~days.
  The coloured banding represents $\pm3$ times the standard-error on
  each median.  Dates indicate January~1 for each year.  There is no
  transmitted-light measured at Las Campanas, and so only the
  scattering values are presented. \label{fig:campanas_extinction}}
\end{figure*}

The BiSON node at Las Campanas is situated~\SI{630}{\kilo\meter} north
of Santiago, at the \emph{Las Campanas Observatory}.  The measured
extinction coefficients over the operational lifetime of the site are
shown in Figure~\ref{fig:campanas_extinction}.

Transmission data from this site are unreliable due to instrumentation
issues, and so only the scattered-light data have been used to derive
the extinction coefficients.  The median extinction is~\num{0.033} for
both morning and afternoon, with standard deviation of less
than~\num{0.02}.  There is also little variation between winter and
summer periods with mean extinction at~\num{0.028} and~\num{0.042}
respectively.  Morning data are collected through airmasses over South
America, and afternoon data are through air over the South Pacific
ocean.  There appears to be no significant difference between the two
zones, and Las Campanas has the most stable atmosphere of all BiSON
sites, showing the lowest standard deviation.

\citet{Siher2002} performed a similar atmospheric analysis for the
IRIS site based at La~Silla, the adjacent mountain ridge to
Las~Campanas.  They quote a value of~\num{0.097}, which is higher than
found here, but we again have to consider the shorter wavelength used
by IRIS, and also that we expect the extinction from the scattered
light to be a slight underestimate of the equivalent extinction from
the transmitted light.  The measured standard deviation is similar
at~\num{0.028}.

\subsection{Narrabri}

\begin{figure*}
\plotone{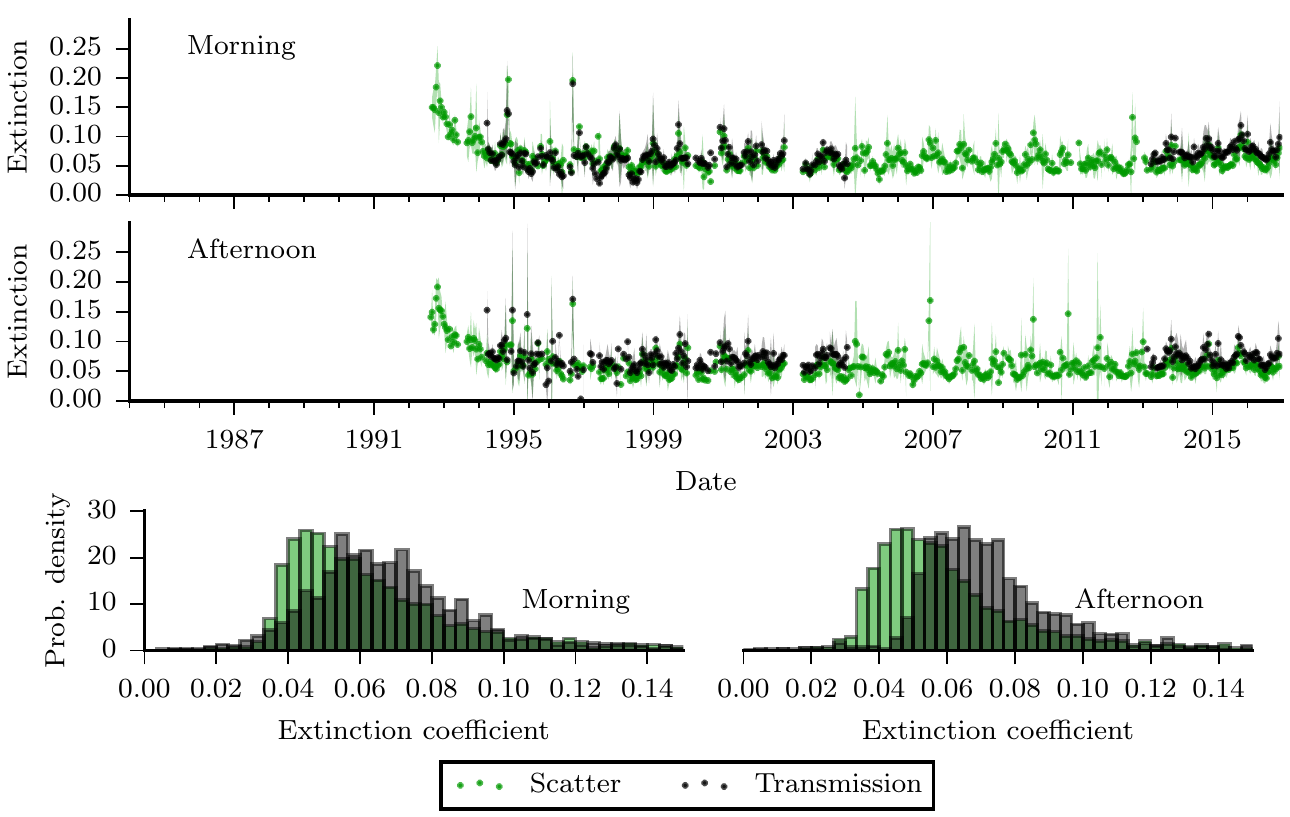}
\caption{Extinction coefficients and statistical distribution from
  Narrabri, eastern Australia.  Each dot represents the median value
  over 14~days.  The coloured banding represents $\pm3$ times the
  standard-error on each median. Dates indicate January~1 for each
  year.\label{fig:narrabri_extinction}}
\end{figure*}

The BiSON node at Narrabri is situated~\SI{525}{\kilo\meter}
north-west of Sydney, at the \emph{Paul Wild Observatory}.  The
measured extinction coefficients over the operational lifetime of the
site are shown in Figure~\ref{fig:narrabri_extinction}.

The median morning extinction is~\num{0.074}, and afternoon
is~\num{0.077}.  The standard deviation is approximately~\num{0.04}.
There are no significant differences in environment around Narrabri
and again this is reflected in the stability of the extinction
coefficients between morning and afternoon.  There is, however, a
strong seasonal variation with median values dropping to~\num{0.06} in
the winter and rising to almost~\num{0.08} in the summer.  Narrabri
tends to have lower extinction and lower standard deviation than the
coastal town of Carnarvon, although it does suffer from a higher
percentage of cloudy days overall.  No datasets could be found for
comparison with Narrabri.

\subsection{Mount Wilson}

\begin{figure*}
\plotone{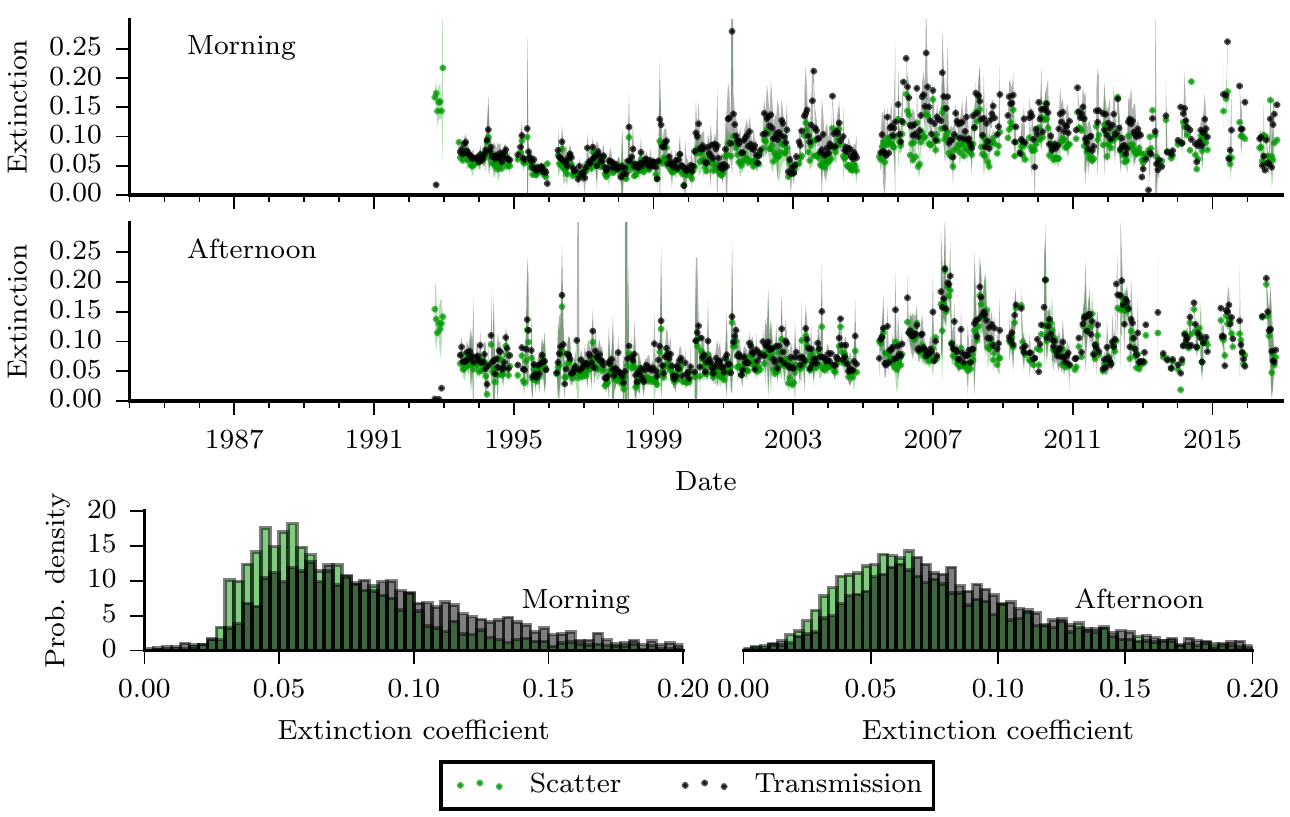}
\caption{Extinction coefficients and statistical distribution from
  Mount Wilson, California USA.  Each dot represents the median value
  over 14~days.  The coloured banding represents $\pm3$ times the
  standard-error on each median. Dates indicate January~1 for each
  year.\label{fig:mtwilson_extinction}}
\end{figure*}
     
The BiSON node at Mount Wilson is situated~\SI{52}{\kilo\meter}
north-east of Los~Angeles, at the \emph{Mount Wilson (Hale)
  Observatory}.  The measured extinction coefficients over the
operational lifetime of the site are shown in
Figure~\ref{fig:mtwilson_extinction}.

The median morning extinction is~\num{0.085}, and afternoon
is~\num{0.081}.  Mount Wilson is an atmospherically-interesting site
due to its location close to the major city of Los~Angeles.  Morning
data are collected over the San~Gabriel mountains, while afternoon
data are observed through air over the city, and so it may be expected
that afternoon extinction values would be worse.  The median-values
found here show that morning and afternoon are generally similar,
however the modal values do show an increase from~\num{0.048} in the
morning to~\num{0.070} in the afternoon.  There is an upward trend in
extinction levels from around 1999, becoming from 2007 part of the
large scatter in values that could be considered to be a result of the
highly variable atmosphere near a large city.  In clear sky conditions
during our observations the inversion boundary layer traps any
pollution at a relatively low altitude --- the well-known Los~Angeles
smog.  Since the observatory elevation is well above the inversion
layer, the pollution would not be seen except at very high zenith
angles that have been specifically excluded from this analysis.  At
this site, light is collected via two mirrors, known as a
c{\oe}lostat, and it is possible that the increase in scatter is due
to gradual reduction in performance due to deterioration of the
mirrors; however, since the extinction is determined from a daily
calibration, any variation in long term performance is removed
completely and so this is just speculation and the cause of the
increased scatter is not clear.  During the winter months, Mount
Wilson does suffer a high proportion of cloudy days, but when the sky
is clear there appears to be little variation between seasons, both in
terms of absolute extinction and standard deviation.

\subsection{Site summary}

A summary of the extinction coefficients from 1995 over all seasons is
given in Table~\ref{table:extinction_coefficients}.  In order to
compare seasonal differences, the values measured over two months for
each mid-summer (July-August in the northern hemisphere,
January-February in the southern hemisphere) are presented in
Table~\ref{table:summer_extinction_coefficients}, and during
mid-winter (January-February in the northern hemisphere, July-August
in the southern hemisphere) in
Table~\ref{table:winter_extinction_coefficients}.

\begin{table*}[p]
    \caption{Extinction coefficients from all sites.  Only 1995
      onwards has been considered in order to remove any exceptional
      atmospheric events.  The data-sets are sufficiently large that
      the standard error of the mean (i.e., $\sigma/\sqrt{N}$) is very
      small and therefore not presented.}
    \label{table:extinction_coefficients}
    \begin{center}
    \begin{tabular*}{\textwidth}{@{\extracolsep\fill}l c c c c c}

    \toprule

    Location & Detector & Mode & Median & Mean & Sigma\\

    \midrule

Mount Wilson &   Transmission (Morning) & 0.048 & 0.085 & 0.096 & 0.055\\
Mount Wilson & Transmission (Afternoon) & 0.070 & 0.081 & 0.093 & 0.057\\
Mount Wilson &        Scatter (Morning) & 0.056 & 0.064 & 0.074 & 0.043\\
Mount Wilson &      Scatter (Afternoon) & 0.061 & 0.070 & 0.082 & 0.053\\
\\
Las Campanas &   Transmission (Morning)\\
Las Campanas & Transmission (Afternoon)\\
Las Campanas &        Scatter (Morning) & 0.029 & 0.033 & 0.037 & 0.018\\
Las Campanas &      Scatter (Afternoon) & 0.028 & 0.033 & 0.036 & 0.014\\
\\
   Iza{\~n}a &   Transmission (Morning) & 0.054 & 0.058 & 0.092 & 0.090\\
   Iza{\~n}a & Transmission (Afternoon) & 0.046 & 0.060 & 0.092 & 0.091\\
   Iza{\~n}a &        Scatter (Morning) & 0.048 & 0.051 & 0.082 & 0.084\\
   Iza{\~n}a &      Scatter (Afternoon) & 0.039 & 0.049 & 0.079 & 0.087\\
\\
  Sutherland &   Transmission (Morning) & 0.038 & 0.046 & 0.054 & 0.040\\
  Sutherland & Transmission (Afternoon) & 0.039 & 0.045 & 0.051 & 0.026\\
  Sutherland &        Scatter (Morning) & 0.034 & 0.039 & 0.048 & 0.039\\
  Sutherland &      Scatter (Afternoon) & 0.037 & 0.038 & 0.044 & 0.027\\
\\
   Carnarvon &   Transmission (Morning) & 0.052 & 0.072 & 0.085 & 0.050\\
   Carnarvon & Transmission (Afternoon) & 0.052 & 0.078 & 0.088 & 0.043\\
   Carnarvon &        Scatter (Morning) & 0.057 & 0.074 & 0.087 & 0.049\\
   Carnarvon &      Scatter (Afternoon) & 0.052 & 0.071 & 0.081 & 0.043\\
\\
    Narrabri &   Transmission (Morning) & 0.053 & 0.066 & 0.070 & 0.029\\
    Narrabri & Transmission (Afternoon) & 0.061 & 0.068 & 0.073 & 0.025\\
    Narrabri &        Scatter (Morning) & 0.048 & 0.055 & 0.062 & 0.028\\
    Narrabri &      Scatter (Afternoon) & 0.050 & 0.053 & 0.059 & 0.027\\

    \botrule

    \end{tabular*}
    \end{center}
\end{table*}

\begin{table*}[p]
  \caption{Summer extinction coefficients from all sites.  In the
    northern hemisphere the measured summer months each year are the
    beginning of July to the end of August, and in the southern
    hemisphere the beginning of January until the end of the February.
    Only 1995 onwards has been considered in order to remove any
    exceptional atmospheric events.  The data-sets are sufficiently
    large that the standard error of the mean (i.e.,
    $\sigma/\sqrt{N}$) is very small and therefore not presented.}
    \label{table:summer_extinction_coefficients}
    \begin{center}
    \begin{tabular*}{\textwidth}{@{\extracolsep\fill}l c c c c c}

    \toprule

    Location & Detector & Mode & Median & Mean & Sigma\\

    \midrule

Mount Wilson &   Transmission (Morning) & 0.053 & 0.077 & 0.084 & 0.041\\
Mount Wilson & Transmission (Afternoon) & 0.050 & 0.082 & 0.095 & 0.062\\
Mount Wilson &        Scatter (Morning) & 0.053 & 0.057 & 0.065 & 0.031\\
Mount Wilson &      Scatter (Afternoon) & 0.037 & 0.076 & 0.088 & 0.056\\
\\
Las Campanas &   Transmission (Morning)\\
Las Campanas & Transmission (Afternoon)\\
Las Campanas &        Scatter (Morning) & 0.032 & 0.042 & 0.046 & 0.017\\
Las Campanas &      Scatter (Afternoon) & 0.040 & 0.041 & 0.044 & 0.016\\
\\
   Iza{\~n}a &   Transmission (Morning) & 0.057 & 0.073 & 0.151 & 0.133\\
   Iza{\~n}a & Transmission (Afternoon) & 0.057 & 0.075 & 0.147 & 0.136\\
   Iza{\~n}a &        Scatter (Morning) & 0.048 & 0.070 & 0.142 & 0.130\\
   Iza{\~n}a &      Scatter (Afternoon) & 0.047 & 0.070 & 0.141 & 0.135\\
\\
  Sutherland &   Transmission (Morning) & 0.043 & 0.046 & 0.050 & 0.022\\
  Sutherland & Transmission (Afternoon) & 0.044 & 0.046 & 0.048 & 0.013\\
  Sutherland &        Scatter (Morning) & 0.036 & 0.040 & 0.044 & 0.021\\
  Sutherland &      Scatter (Afternoon) & 0.037 & 0.040 & 0.042 & 0.014\\
\\
   Carnarvon &   Transmission (Morning) & 0.085 & 0.100 & 0.115 & 0.078\\
   Carnarvon & Transmission (Afternoon) & 0.095 & 0.107 & 0.114 & 0.045\\
   Carnarvon &        Scatter (Morning) & 0.119 & 0.106 & 0.119 & 0.068\\
   Carnarvon &      Scatter (Afternoon) & 0.081 & 0.103 & 0.110 & 0.046\\
\\
    Narrabri &   Transmission (Morning) & 0.076 & 0.074 & 0.081 & 0.030\\
    Narrabri & Transmission (Afternoon) & 0.073 & 0.077 & 0.086 & 0.041\\
    Narrabri &        Scatter (Morning) & 0.060 & 0.068 & 0.074 & 0.026\\
    Narrabri &      Scatter (Afternoon) & 0.060 & 0.061 & 0.068 & 0.035\\
    
    \botrule

    \end{tabular*}
    \end{center}
\end{table*}

\begin{table*}[p]
  \caption{Winter extinction coefficients from all sites.  In the
    northern hemisphere the measured winter months each year are the
    beginning of January until the end of the February, and in the
    southern hemisphere the beginning of July to the end of August.
    Only 1995 onwards has been considered in order to remove any
    exceptional atmospheric events.  The data-sets are sufficiently
    large that the standard error of the mean (i.e.,
    $\sigma/\sqrt{N}$) is very small and therefore not presented.}
    \label{table:winter_extinction_coefficients}
    \begin{center}
    \begin{tabular*}{\textwidth}{@{\extracolsep\fill}l c c c c c}

    \toprule

    Location & Detector & Mode & Median & Mean & Sigma\\

    \midrule

Mount Wilson &   Transmission (Morning) & 0.048 & 0.077 & 0.086 & 0.051\\
Mount Wilson & Transmission (Afternoon) & 0.075 & 0.070 & 0.074 & 0.032\\
Mount Wilson &        Scatter (Morning) & 0.046 & 0.060 & 0.070 & 0.035\\
Mount Wilson &      Scatter (Afternoon) & 0.053 & 0.056 & 0.059 & 0.027\\
\\
Las Campanas &   Transmission (Morning)\\
Las Campanas & Transmission (Afternoon)\\
Las Campanas &        Scatter (Morning) & 0.028 & 0.028 & 0.031 & 0.015\\
Las Campanas &      Scatter (Afternoon) & 0.027 & 0.028 & 0.030 & 0.010\\
\\
   Iza{\~n}a &   Transmission (Morning) & 0.045 & 0.048 & 0.059 & 0.035\\
   Iza{\~n}a & Transmission (Afternoon) & 0.045 & 0.048 & 0.062 & 0.062\\
   Iza{\~n}a &        Scatter (Morning) & 0.042 & 0.043 & 0.051 & 0.030\\
   Iza{\~n}a &      Scatter (Afternoon) & 0.038 & 0.038 & 0.047 & 0.045\\
\\
  Sutherland &   Transmission (Morning) & 0.038 & 0.049 & 0.061 & 0.042\\
  Sutherland & Transmission (Afternoon) & 0.039 & 0.044 & 0.056 & 0.046\\
  Sutherland &        Scatter (Morning) & 0.033 & 0.041 & 0.057 & 0.056\\
  Sutherland &      Scatter (Afternoon) & 0.029 & 0.037 & 0.049 & 0.045\\
\\
   Carnarvon &   Transmission (Morning) & 0.051 & 0.056 & 0.065 & 0.028\\
   Carnarvon & Transmission (Afternoon) & 0.056 & 0.060 & 0.065 & 0.026\\
   Carnarvon &        Scatter (Morning) & 0.051 & 0.060 & 0.068 & 0.032\\
   Carnarvon &      Scatter (Afternoon) & 0.052 & 0.053 & 0.059 & 0.026\\
\\
    Narrabri &   Transmission (Morning) & 0.057 & 0.058 & 0.061 & 0.024\\
    Narrabri & Transmission (Afternoon) & 0.064 & 0.061 & 0.064 & 0.016\\
    Narrabri &        Scatter (Morning) & 0.040 & 0.046 & 0.050 & 0.017\\
    Narrabri &      Scatter (Afternoon) & 0.044 & 0.046 & 0.049 & 0.016\\

    \botrule

    \end{tabular*}
    \end{center}
\end{table*}

%
%
%
%

\section{Conclusion}
\label{S-Conclusion}

Over 30 years of helioseismic data have been acquired by BiSON from
several international observatories, the locations of which are
summarised in Table~\ref{table:bisoncoordinates}.  In this paper we
have made innovative use of these data to derive measurements of
atmospheric opacity in the $\mathrm{I_c}$ band, and we have presented
the column atmospheric extinction coefficients from each site over the
years 1984 to 2016.  This is an important contribution to the
literature since there are limited data on aerosol optical depth from
other sources prior to the mid-1990s.

\begin{figure*}
\plotone{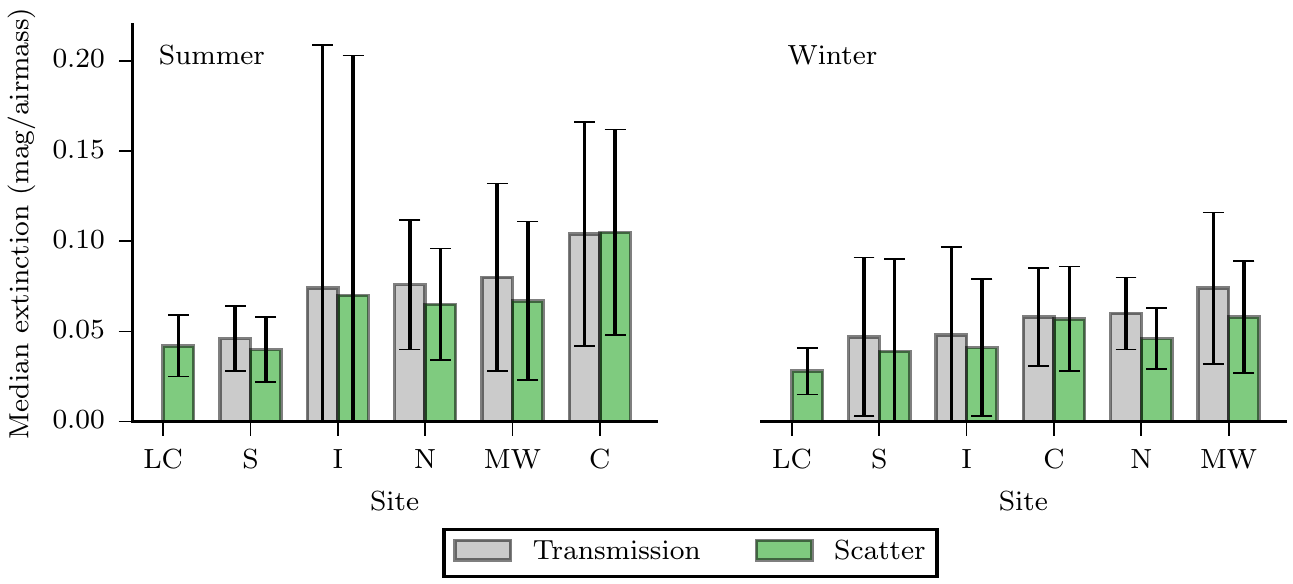}
\caption{Summary of median extinction coefficients and distribution
  standard deviations.  Only 1995 onwards has been considered in order
  to remove any exceptional atmospheric events.  Left panel: Summer
  periods from Table~\ref{table:summer_extinction_coefficients}.
  Right panel: Winter periods from
  Table~\ref{table:winter_extinction_coefficients}.  The mean of the
  morning and afternoon median coefficients have been taken for each
  site, and the extinction distribution standard deviation shown as
  error bars.  The x-axis labels are the initials of each site
  name. \label{fig:site_comparison}}
\end{figure*}

The median and standard deviation values for mid-summer periods from
Table~\ref{table:summer_extinction_coefficients}, and for mid-winter
periods from Table~\ref{table:winter_extinction_coefficients}, are
shown graphically in Figure~\ref{fig:site_comparison}.  The mean of
the morning and afternoon values were taken for each site, and the
extinction distribution standard deviation shown as error bars.  We
find the best results are from the Las Campanas and Sutherland
observatories with consistent year-round performance.  Iza{\~n}a
offers comparable high-performance during the winter months, but
becomes the worst site during mid-summer due to high concentrations of
mineral dust in the atmosphere from the Western Sahara on the North
African coast.  Carnarvon similarly suffers degradation during the
summer months, which again is most likely due to wind-borne sand in
the dry enviroment of northern Western Australia.  Narrabri and Mount
Wilson both offer consistent performance.

The atmospheric extinction data presented here and the code to
generate the figures are open access and can be downloaded from the
University of Birmingham ePapers data archive~\citep{epapers3022}.

\acknowledgments

We would like to thank all those who have been associated with {BiSON}
over the years.  We particularly acknowledge the technical assistance
at our remote network sites, with sincere apologies to anyone
inadvertently missed:
In Mount Wilson: Stephen Pinkerton, the team of {USC} undergraduate
observing assistants, former {USC} staff members Maynard Clark, Perry
Rose, Natasha Johnson, Steve Padilla, and Shawn Irish, and former
{UCLA} staff members Larry Webster and John Boyden.
In Las Campanas: Patricio Pinto, Andres Fuentevilla, Emilio Cerda,
Frank Perez, Marc Hellebaut, Patricio Jones, Gast{\'o}n Gutierrez,
Juan Navarro, Francesco Di Mille, Roberto Bermudez, and the staff of
{LCO}.
In Iza{\~n}a: All staff at the {IAC} who have contributed to running
the {Mark~I} instrument over many years.
In Sutherland: Pieter Fourie, Willie Koorts, Jaci Cloete, Reginald
Klein, John Stoffels, and the staff of {SAAO}.
In Carnarvon: Les Bateman, Les Schultz, Sabrina Dowling-Giudici, Inge
Lauw of Williams and Hughes Lawyers, and {NBN} Co.\ Ltd.
In Narrabri: Mike Hill and the staff of {CSIRO}.
MNL acknowledges the support of The Danish Council for Independent
Research~|~Natural Sciences (Grant DFF-4181-00415).  Funding for the
Stellar Astrophysics Centre (SAC) is provided by The Danish National
Research Foundation (Grant DNRF106).  {BiSON} is funded by the Science
and Technology Facilities Council ({STFC}).  We thank the anonymous
reviewers for their help and useful comments.

%






\appendix

%
%
%
%

\section{Details on {BiSON} observations and line-of-sight velocity effects}
\label{S-Line-of-Sight}

The BiSON solar spectrometers precisely measure the line-of-sight
velocity of the solar surface by looking at the Doppler shift of a
Fraunhofer line using a potassium vapour reference-cell.

\begin{figure}
\plottwo{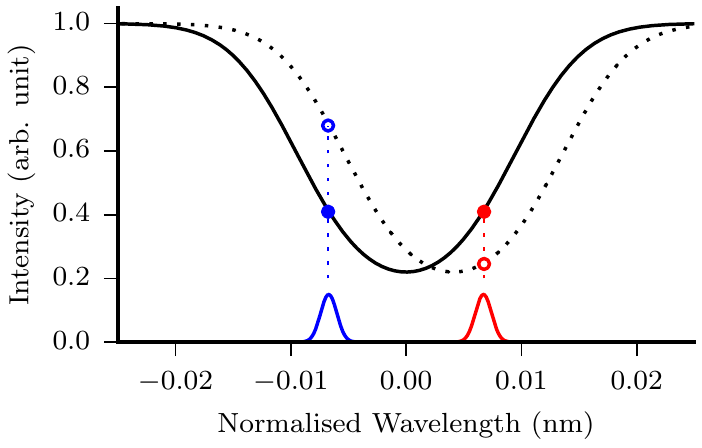}{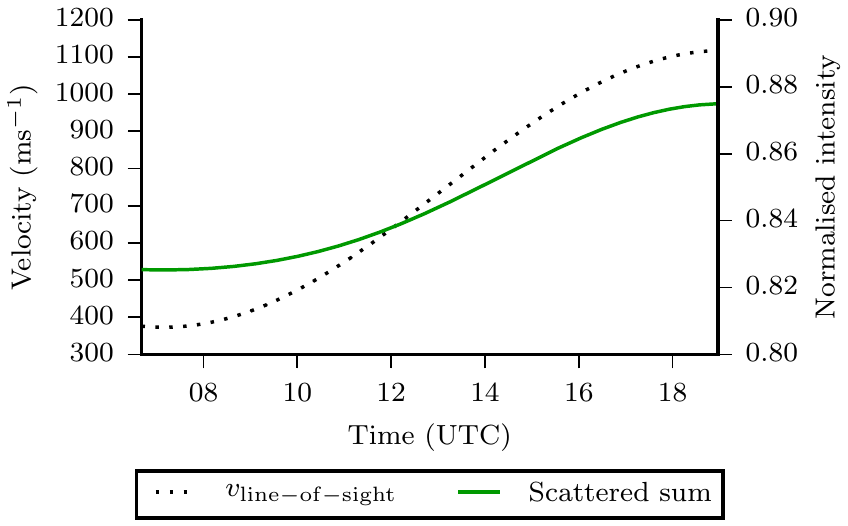}
\caption{Left: The solid line represents an absorption line with zero
  line-of-sight velocity offset.  The sum of the intensity measurement
  of the two wings (solid circles) is 0.82.  The dashed line
  represents an offset where $v_\mathrm{line-of-sight}$ is
  \SI{1600}{\metre\per\second}.  The sum of the intensity measurement
  of the two wings (open circles) rises to 0.93.  Neither measurement
  returns the unity intensity measured outside the absorption
  line. Right: The expected variation in scattered-sum over typical
  line-of-sight Doppler velocities, in the absence of any atmospheric
  extinction.  The amplitude of the daily change in
  effective-intensity varies throughout the year with changes in
  Earth's orbital velocity.  \label{fig:line-of-sight}}
\end{figure}

The line-of-sight velocity is dominated by three components: the
rotation of the Earth, the orbital velocity of the Earth, and the
gravitational red-shift.  At high air-masses the extended source of
the Sun also suffers differential extinction which causes the
line-of-sight velocity due to solar rotation to become unequally
weighted across the solar disc~\citep{doi:10.1093/mnras/stu803}.  In
addition to these factors are small oscillations of the solar surface
which are the primary science output of the instrument, but for this
analysis can be considered to be insignificant.  The lab-frame
absorption line is Zeeman-split by placing the vapour-cell in a
longitudinal magnetic field which, when combined with suitable
polarisation control, places the passbands of the spectrometer on the
wings of the corresponding solar absorption-line at the points that
provide the highest sensitivity to Doppler-shift caused by the
line-of-sight velocity.  By taking the sum of the intensity measured
on both wings of the absorption line, this can be used as a proxy for
the total intensity.  If the solar absorption line is modelled as a
thermally and rotationally broadened Gaussian, then the left panel of
Figure~\ref{fig:line-of-sight} shows the expected scattered-sum of the
two wings when the Doppler-velocity is zero, and for when the
Doppler-velocity is at the maximum expected red-shift
of~\SI{1600}{\metre\per\second}.  Neither scattered-sum correctly
recovers the unity-intensity continuum outside the Gaussian that would
be measured from the transmitted light.

The difference between the measurement of the transmitted light and
the scattered-sum varies with Doppler-velocity.  The right panel of
Figure~\ref{fig:line-of-sight} shows the expected effect of this in
the absence of atmospheric extinction over the range of line-of-sight
velocities experienced during a typical day.  The amplitude of the
daily change in effective-intensity varies throughout the year with
changes in Earth's orbital velocity, and this creates a very small
seasonal effect in the scattering extinction coefficients.  The
increased effective-intensity is most pronounced at high red-shifts
experienced during the local-afternoon of each site.  The gradual
increase in effective-intensity will cause any extinction coefficients
derived from the scattered-sum to have a slight over-estimate in the
morning and a larger under-estimate in the afternoon.

If the precise shape and depth of the solar absorption-line were
known, then it would be possible to correct for this effect and
recover the equivalent transmitted-intensity.  Unfortunately the solar
disc-integrated line profile changes with magnetic activity and so one
can not correct based on a single line
profile.  \citet{2014AMT.....7.4103B} have resolved this problem for
our data from Iza{\~n}a by making use of a quasi-continuous Langley
calibration technique, where they have achieved a \emph{mean bias}
(defined as the mean difference between the transmission and
scattering extinction coefficients) of $\le 0.01$.

In practice, extinction data from the scattering detectors offers a
slight under-estimate of extinction in comparison to transmitted
measurements for both morning and afternoon periods.  However, this is
simply a small systematic offset on the absolute value, and the
extinction coefficients observed from real data are otherwise
identical to those derived from the transmitted light and show the
same trends.




\bibliography{references}

\begin{thebibliography}{}
\expandafter\ifx\csname natexlab\endcsname\relax\def\natexlab#1{#1}\fi
\providecommand{\url}[1]{\href{#1}{#1}}

\bibitem[{{Barreto} {et~al.}(2014){Barreto}, {Cuevas}, {Pall{\'e}}, {Romero},
  {Guirado}, {Wehrli}, \& {Almansa}}]{2014AMT.....7.4103B}
{Barreto}, A., {Cuevas}, E., {Pall{\'e}}, P., {et~al.} 2014, Atmospheric
  Measurement Techniques, 7, 4103

\bibitem[{Davies {et~al.}(2014)Davies, Chaplin, Elsworth, \&
  Hale}]{doi:10.1093/mnras/stu803}
Davies, G.~R., Chaplin, W.~J., Elsworth, Y., \& Hale, S.~J. 2014, Monthly
  Notices of the Royal Astronomical Society, 441, 3009.
\newblock \url{http://dx.doi.org/10.1093/mnras/stu803}

\bibitem[{{Fischer} {et~al.}(1986){Fischer}, {Hill}, {Jones}, {Leibacher},
  {McCurnin}, {Stebbins}, \& {Wagner}}]{1986SoPh..103...33F}
{Fischer}, G., {Hill}, F., {Jones}, W., {et~al.} 1986, \solphys, 103, 33

\bibitem[{Garc{\'i}a-Gil {et~al.}(2010)Garc{\'i}a-Gil, Mu{\~n}oz-Tu{\~n}{\'o}n,
  \& Varela}]{1538-3873-122-895-1109}
Garc{\'i}a-Gil, A., Mu{\~n}oz-Tu{\~n}{\'o}n, C., \& Varela, A.~M. 2010,
  Publications of the Astronomical Society of the Pacific, 122, 1109.
\newblock \url{http://stacks.iop.org/1538-3873/122/i=895/a=1109}

\bibitem[{{Guerrero} {et~al.}(1998){Guerrero}, {Garc{\'{\i}}a-L{\'o}pez},
  {Corradi}, {Jim{\'e}nez}, {Fuensalida}, {Rodr{\'{\i}}guez-Espinosa},
  {Alonso}, {Centuri{\'o}n}, \& {Prada}}]{1998NewAR..42..529G}
{Guerrero}, M.~A., {Garc{\'{\i}}a-L{\'o}pez}, R.~J., {Corradi}, R.~L.~M.,
  {et~al.} 1998, \nar, 42, 529

\bibitem[{Hale(2017)}]{epapers3022}
Hale, S.~J. 2017, {BiSON - Atmospheric extinction coefficients in the
  $\mathrm{I_c}$ band - 1984 to 2016},  University of Birmingham, UK:
  Birmingham Solar Oscillations Network.
\newblock \url{http://epapers.bham.ac.uk/3022/}

\bibitem[{Hale {et~al.}(2016)Hale, Howe, Chaplin, Davies, \&
  Elsworth}]{Hale2016}
Hale, S.~J., Howe, R., Chaplin, W.~J., Davies, G.~R., \& Elsworth, Y.~P. 2016,
  Solar Physics, 291, 1.
\newblock \url{http://dx.doi.org/10.1007/s11207-015-0810-0}

\bibitem[{{Hill} {et~al.}(1994){Hill}, {Fischer}, {Forgach}, {Grier},
  {Leibacher}, {Jones}, {Jones}, {Kupke}, {Stebbins}, \&
  {Clay}}]{1994SoPh..152..351H}
{Hill}, F., {Fischer}, G., {Forgach}, S., {et~al.} 1994, \solphys, 152, 351

\bibitem[{{Jimenez} {et~al.}(1998){Jimenez}, {Gonzalez Jorge}, \&
  {Rabello-Soares}}]{1998A&AS..129..413J}
{Jimenez}, A., {Gonzalez Jorge}, H., \& {Rabello-Soares}, M.~C. 1998, \aaps,
  129, 413

\bibitem[{Kasten \& Young(1989)}]{Kasten:89}
Kasten, F., \& Young, A.~T. 1989, Appl. Opt., 28, 4735.
\newblock \url{http://ao.osa.org/abstract.cfm?URI=ao-28-22-4735}

\bibitem[{{Kilkenny}(1995)}]{1995Obs...115...25K}
{Kilkenny}, D. 1995, The Observatory, 115, 25

\bibitem[{{King}(1985)}]{tn031}
{King}, D.~L. 1985, {Atmospheric Extinction at the Roque de los Muchachos
  Observatory, La Palma}, {RGO/La Palma Technical Note}~31, {Instituto de
  Astrof{\`i}sica de Canarias}, La Laguna, Tenerife, Canary Islands.
\newblock
  \url{http://www.ing.iac.es/Astronomy/observing/manuals/ps/tech_notes/tn031.pdf}

\bibitem[{Laken {et~al.}(2016)Laken, Parviainen, Garc{\'i}a-Gil,
  Mu{\~n}oz-Tu{\~n}{\'o}n, Varela, Fernandez-Acosta, \&
  {Pall{\'e}}}]{doi:10.1175/JCLI-D-14-00600.1}
Laken, B.~A., Parviainen, H., Garc{\'i}a-Gil, A., {et~al.} 2016, Journal of
  Climate, 29, 227

\bibitem[{Laken {et~al.}(2014)Laken, Parviainen, {Pall{\'e}}, \&
  Shahbaz}]{QJ:QJ2170}
Laken, B.~A., Parviainen, H., {Pall{\'e}}, E., \& Shahbaz, T. 2014, Quarterly
  Journal of the Royal Meteorological Society, 140, 1058.
\newblock \url{http://dx.doi.org/10.1002/qj.2170}

\bibitem[{{Roca Cort{\'e}s} \& {Pall{\'e}}(2014)}]{2014MNRAS.443.1837R}
{Roca Cort{\'e}s}, T., \& {Pall{\'e}}, P.~L. 2014, \mnras, 443, 1837

\bibitem[{Siher {et~al.}(2002)Siher, Benkhaldoun, \& Fossat}]{Siher2002}
Siher, E., Benkhaldoun, Z., \& Fossat, E. 2002, Experimental Astronomy, 13,
  159.
\newblock \url{http://dx.doi.org/10.1023/A:1025535615069}

\bibitem[{{Siher} {et~al.}(2004){Siher}, {Ortolani}, {Sarazin}, \&
  {Benkhaldoun}}]{2004SPIE.5489..138S}
{Siher}, E.~A., {Ortolani}, S., {Sarazin}, M.~S., \& {Benkhaldoun}, Z. 2004, in
  \procspie, Vol. 5489, Ground-based Telescopes, ed. J.~M. {Oschmann}, Jr.,
  138--145

\end{thebibliography}






\end{document}